\documentclass[aps,prd,twocolumn,showpacs,superscriptaddress,groupedaddress,nofootinbib]{revtex4-2}  % for review and submission
\usepackage{graphicx}  % needed for figures
\usepackage{dcolumn}   % needed for some tables
\usepackage{bm}        % for math
\usepackage{amssymb}   % for math
\usepackage{amsmath}
\usepackage{slashed}
\usepackage[unicode=true,
 bookmarks=true,bookmarksnumbered=false,bookmarksopen=false,
 breaklinks=false,pdfborder={0 0 1},backref=false,colorlinks=true]{hyperref}
\usepackage{cleveref}
\usepackage{tikz-feynman}

\hypersetup{citecolor=blue,linkcolor=blue,urlcolor=blue}

\allowdisplaybreaks

\def\fSC#1#2#3{ f_{#1 #2}^{\;\;\;\; #3} }

\begin{document}

\hfill CERN-TH-2024-086
% the following line is for submission, including submission to the arXiv!!
%\hspace{5.2in} \mbox{Fermilab-Pub-04/xxx-E}

\title{A Double Copy of Geometry}
%\input author_list.tex       % D0 authors (remove the first 3 lines
                             % of this file prior to submission, they
			     % contain a time stamp for the authorlist)
\author{Andreas Helset}                             % (includes institutions and visitors)
\affiliation{ Theoretical Physics Department, CERN, 1211 Geneva 23, Switzerland }
\date{\today}

\begin{abstract}
        We propose an extension of the color-kinematics duality for nonlinear sigma models with nonsymmetric cosets. 
        The role of color is played by curvatures in field space. The duality between curvature and kinematics 
        allows for a new double copy: the double copy of the nonlinear sigma model for nonsymmetric cosets is the general Galileon theory. We provide evidence for this new double copy of geometry.
\end{abstract}

\maketitle

%%%%%%%%%%%%%%%%%%%%%%%%%%%%%%%%%%%%%%%%%%%%%%%%%%%%%%%
%%%%%%%%%%%%%%%% Begin Intro %%%%%%%%%%%%%%%%%%%%%%%%%%
%%%%%%%%%%%%%%%%%%%%%%%%%%%%%%%%%%%%%%%%%%%%%%%%%%%%%%%

\section{\label{sec:intro}Introduction}

The duality between color and kinematics hints at a rich connection between apparently disparate theories. Through this duality, various theories can be recast as double copies of other theories \cite{Bern:2008qj,Bern:2010ue}. The double-copy relation was first discovered for open and closed string amplitudes \cite{Kawai:1985xq} and---in the field-theory limit---links gauge theory and gravity amplitudes. Since then, a web of theories is connected through the double copy (see refs.~\cite{Bern:2019prr,Bern:2022wqg} for recent reviews).

Of special interest to us is the double copy between scalar effective field theories. The prime example is the nonlinear sigma model (NLSM) for symmetric cosets, whose double copy results in amplitudes for the special Galileon theory. Both of these theories are exceptional. The soft limit of the NLSM for a symmetric coset vanishes, known as the Adler zero \cite{Adler:1964um}. For the special Galileon theory, the soft limit also vanishes, but at a higher rate than naively expected \cite{Cheung:2014dqa,Hinterbichler:2015pqa,Cheung:2016drk}. These conditions are sufficient to fully reconstruct higher-point amplitudes from lower-point amplitudes in these theories through on-shell recursion relations.

Close cousins of these theories also exhibit special features. The NLSM for a nonsymmetric coset has a nonvanishing soft limit. However, this limit is still universal. The soft theorem is \cite{Cheung:2021yog,Derda:2024jvo}
\begin{align}
	\lim_{q\rightarrow 0} \mathcal{A}_{n+1} = \nabla_{i} \mathcal{A}_{n} ,
\end{align}
which holds for any effective field theory with a massless scalar but without a potential. Here, $\mathcal{A}_{n}$ is the $n$-point amplitude, $\nabla_{i}$ is the covariant derivative in field space, and $i$ is the flavor label of the soft particle with momentum $q$. As for the general Galileon theory, its soft limit vanishes, albeit not as fast as for the special Galileon theory. 

In this paper, we explore whether these scalar effective field theories can be included under the umbrella of color-kinematics dual theories. We will present evidence through an ansatz approach that the answer is affirmative. As we will see, the rules of the game have to be slightly modified to accommodate these more general theories. However, the underlying philosophy of finding a duality between color and kinematics prevails.

%%%%%%%%%%%%%%%%%%%%%%%%%%%%%%%%%%%%%%%%%%%%%%%%%%%%%%%
%%%%%%%%%%%%%%%% End Intro %%%%%%%%%%%%%%%%%%%%%%%%%%%%
%%%%%%%%%%%%%%%%%%%%%%%%%%%%%%%%%%%%%%%%%%%%%%%%%%%%%%%

%%%%%%%%%%%%%%%%%%%%%%%%%%%%%%%%%%%%%%%%%%%%%%%%%%%%%%%
%%%%%%%%%%%%%%%% Begin Geometry of Field Space %%%%%%%%
%%%%%%%%%%%%%%%%%%%%%%%%%%%%%%%%%%%%%%%%%%%%%%%%%%%%%%%

\section{\label{sec:Geometry}The Geometry of the NLSM}

The NLSM describes the leading interactions for Goldstone bosons that arise from the spontaneous breaking of a global symmetry $G$ down to a subgroup $H$. The Lagrangian takes the form
\begin{align}
	\label{eq:Lagr1NLSM}
	\mathcal{L}_{\rm NLSM} = \frac{1}{2} g_{ij}(\pi) (\partial_{\mu} \pi)^{i} (\partial^{\mu} \pi)^{j} ,
\end{align}
where $g_{ij}(\pi)$ is the metric on the coset manifold. To ensure that the Lagrangian is invariant under the global symmetry, we employ the CCWZ construction \cite{Coleman:1969sm,Callan:1969sn}. The generators of the Lie algebra are divided into broken $X_{i}$ and unbroken $T_{a}$ generators. The Lie algebra is
\begin{align}
	[T_a, T_b] &= i \fSC{a}{b}{c} T_{c} , \\
	[T_a, X_i] &= i \fSC{a}{i}{j} X_{j} , \\
	[X_i, X_j] &= i \fSC{i}{j}{c} T_{c} + i \fSC{i}{j}{k} X_{k} .
\end{align}
We assume that the coset is reductive and that the structure constants are completely antisymmetric, as for compact groups $G$. If we further assumed that the coset was symmetric, we would set $\fSC{i}{j}{k} = 0$. Crucially, we will not make this restriction and instead keep the additional structure constant. 

Using the exponential map $\xi(x) = e^{i \pi^j(x) X_{j}/F_\pi}$, we construct
\begin{align}
	i (D_{\mu} \pi)^{j} X_j/F_\pi = \xi^{-1} \partial_{\mu} \xi\vert_{X} = e^{i}_{\; k}(\pi) i (\partial_{\mu} \pi)^{k} X_j/F_\pi 
\end{align}
where $e^{i}_{\; k}(\pi)$ is a vierbein on the coset manifold \cite{Alonso:2016oah}. From this, we build an invariant Lagrangian
\begin{align}
	\label{eq:Lagr2NLSM}
	\mathcal{L}_{\rm NLSM} = \frac{1}{2} \delta_{ij} (D_{\mu} \pi)^{i} (D_{\mu} \pi)^{j} .
\end{align}
By equating \cref{eq:Lagr1NLSM,eq:Lagr2NLSM}, we can evaluate the metric $g_{ij}$ in terms of the structure constants of the Lie algebra. The Riemann curvature, which can be obtained from the metric, is
\begin{align}
	R_{ijkl} = f_{ij}^{\;\;\; a} f_{kl a} - \frac{1}{4} f_{ij}^{\;\;\;\; n} f_{kln} , 
\end{align}
while the covariant derivative of the curvature is
\begin{align}
	\label{eq:DR}
	\nabla_{k} R_{ijlm} = \frac{1}{2} f_{ij}^{\;\;\; a} f_{a k}^{\;\;\; n} f_{n lm} - \frac{1}{2} f_{ij}^{\;\;\; n} f_{n k}^{\;\;\; a} f_{a lm} ,
\end{align}
evaluated at the vacuum expectation value (VEV) of the scalar fields. 
Higher curvature terms can be calculated straightforwardly. Note that curvature terms of the form $\nabla^{n} R$ depend on the structure constant $f_{ijk}$. An immediate consequence is that for a symmetric space, where $f_{ijk} = 0$, these curvature terms will vanish and odd-point amplitudes will be zero.

Scattering amplitudes can be expressed as combinations of kinematics and curvatures. Examples for 4 and 5 particles are
\begin{align}
	\label{eq:NLSM4pt}
	\mathcal{A}_{4} =& R_{ijkl} s_{ik} + R_{ikjl} s_{ij} ,  \\
	\label{eq:NLSM5pt}
	\mathcal{A}_{5} =& \nabla_{k} R_{iljm} s_{lm} + \nabla_{l} R_{ikjm} s_{km} + \nabla_{l} R_{ijkm} s_{jm} 
	\nonumber \\ &
	+\nabla_{m} R_{ikjl} s_{kl} +\nabla_{m} R_{ijkl} (s_{jl} + s_{lm}) ,
\end{align}
where $s_{ij} = (p_i + p_j)^2$.

%%%%%%%%%%%%%%%%%%%%%%%%%%%%%%%%%%%%%%%%%%%%%%%%%%%%%%%
%%%%%%%%%%%%%%%% End Geometry of Field Space %%%%%%%%%%
%%%%%%%%%%%%%%%%%%%%%%%%%%%%%%%%%%%%%%%%%%%%%%%%%%%%%%%

%%%%%%%%%%%%%%%%%%%%%%%%%%%%%%%%%%%%%%%%%%%%%%%%%%%%%%%
%%%%%%%%%%%%%%%% Begin General Galileon %%%%
%%%%%%%%%%%%%%%%%%%%%%%%%%%%%%%%%%%%%%%%%%%%%%%%%%%%%%%

\section{\label{sec:Galileon}General Galileon theory}

The other theory we need is the general Galileon theory. The Lagrangian is \cite{Cheung:2016drk}
\begin{align}
	\mathcal{L}_{\rm Gal} &= \sum_{n=1}^{d+1} d_{n} \phi \mathcal{L}_{n-1}^{\rm der}  , \\
	\mathcal{L}_{n-1}^{\rm der} &= (-1)^{d-1} (d-n)! {\rm det} \left\{ \partial^{\nu_i}_{\;\; \nu_j} \phi \right\}^{n}_{i,j=1} .
\end{align}
It has a Galilean shift symmetry $\phi \rightarrow \phi + a + b\cdot x$. As a consequence, it vanishes as $\mathcal{O}(q^2)$ in the soft limit where $q$ is the soft momentum. Due to a Gram-determinant constraint, the theory contains only $d+1$ terms in $d$ dimensions. We will work in general dimensions to not having to take this constraint into account.

It is possible to enforce a stronger shift symmetry \cite{Hinterbichler:2015pqa}. In doing so, all coefficients are uniquely fixed. The resulting theory is the special Galileon amplitude. All odd-point amplitudes are zero and the soft limit is enhanced to $\mathcal{O}(q^3)$.

%%%%%%%%%%%%%%%%%%%%%%%%%%%%%%%%%%%%%%%%%%%%%%%%%%%%%%%
%%%%%%%%%%%%%%%% End General Galileon %%%%
%%%%%%%%%%%%%%%%%%%%%%%%%%%%%%%%%%%%%%%%%%%%%%%%%%%%%%%

%%%%%%%%%%%%%%%%%%%%%%%%%%%%%%%%%%%%%%%%%%%%%%%%%%%%%%%
%%%%%%%%%%%%%%%% Begin Geometry-Kinematics Duality %%%%
%%%%%%%%%%%%%%%%%%%%%%%%%%%%%%%%%%%%%%%%%%%%%%%%%%%%%%%

\section{\label{sec:CKduality}Curvature and kinematics}

To find a color-kinematics dual representation of an $n$-point scattering amplitude, we start by writing the amplitude in terms of a sum over trivalent graphs, $\Gamma$,
\begin{align}
	\label{eq:trivalent}
	\mathcal{A}_{n} = \sum_{i\in \Gamma} \frac{ c_{i} n_{i} }{d_{i}} .
\end{align}
Here, $c_{i}$ is the color factor, $n_{i}$ is the kinematic numerator, and $d_{i}$ is the product of propagators for the corresponding graph. 
Typically, this representation of the amplitude is not unique because the color structures satisfy various relations, such as the Jacobi identity
\begin{align}
	c_i + c_j = c_k .
\end{align}
We can use this freedom to rearrange the kinematic numerators. If we find a representation where the kinematic numerators satisfy the same algebraic relations as the color factor, we declare it a color-kinematics dual representation. For instance, the kinematic numerators should satisfy
\begin{align}
	n_i + n_j = n_k .
\end{align}
The color-kinematics duality is most immediate when the particles are in the adjoint representation. The color factors will then be products of structure constants, $f_{abc}$. 

The magic of the color-kinematics duality happens when we consider replacing the color factor by a kinematic factor. The resulting amplitude is
\begin{align}
	\mathcal{M}_{n} = \sum_{i\in\Gamma} \frac{ n_{i} n_{i}^{\prime} }{d_{i}} ,
\end{align}
where we allow for the possibility that the two kinematic numerators are different (e.g. coming from different theories).

We will search for representations of the NLSM for nonsymmetric cosets that satisfy this duality between color and kinematics. The role of the color factors will be played by the curvature terms in the amplitudes. We call this a {\it curvature-kinematics duality},\footnote{Not to be confused with the geometry-kinematics duality \cite{Cheung:2022vnd}, which pertains to effective operators and field redefinitions with more derivatives.} where $c_i$ now are viewed as the curvature terms. There will be new curvature terms as we go from $n$-point to $(n+1)$-point amplitudes. To ease our search for a dual representation, we will restrict our attention to terms with at most one factor of $f_{ijk}$. In practice, this will allow $\nabla R$ to enter, but $\nabla^n R$ with $n\geq 2$ will be set to zero.

%%%%%%%%%%%%%%%%%%%%%%%%%%%%%%%%%%%%%%%%%%%%%%%%%%%%%%%
%%%%%%%%%%%%%%%% End Geometry-Kinematics Duality %%%%%%
%%%%%%%%%%%%%%%%%%%%%%%%%%%%%%%%%%%%%%%%%%%%%%%%%%%%%%%

%%%%%%%%%%%%%%%%%%%%%%%%%%%%%%%%%%%%%%%%%%%%%%%%%%%%%%%
%%%%%%%%%%%%%%%% Begin Examples %%%%%%%%%%%%%%%%%%%%%%%
%%%%%%%%%%%%%%%%%%%%%%%%%%%%%%%%%%%%%%%%%%%%%%%%%%%%%%%

%
\subsection*{4-point amplitudes}

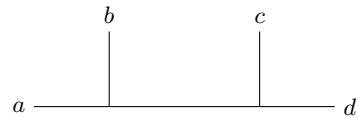
\begin{figure}
\centering
\begin{tikzpicture}
  \draw (-2.0,0) node[left] {$a$} -- (2.0,0) node[right] {$d$};
  \draw (-1.0,0) node[left] {} -- (-1.0,1) node[above] {$b$};
  \draw (1.0,0) node[above] {} -- (1.0,1) node[above] {$c$};
\end{tikzpicture}
\caption{Trivalent graph for four particles.}
\label{fig:graph4pt}
\end{figure}

As a primer on the method we will employ at higher points, we start with the four-point amplitude. 
The graph is shown in \cref{fig:graph4pt}, and the corresponding propagator and curvature factor are 
\begin{align}
	d(a,b,c,d) &= s_{ab} , \\
	c(a,b,c,d) &= R_{abcd} .
\end{align}
The usual adjoint color factor $f_{abx}f_{xcd}$ and the Riemann curvature tensor $R_{abcd}$ satisfy all the same identities at four points; the Riemann curvature tensor has the same symmetry properties as the adjoint color factor, and the Bianchi identity for the curvature serves the same role as the Jacobi identity for the color factors:
\begin{align}
	R_{abcd} + &R_{acdb} + R_{adbc} = 0 \qquad 
	\nonumber \\ 
	&\leftrightarrow  \qquad
	\nonumber \\ 
	f_{abx}f_{xcd} + &f_{acx}f_{xdb} + f_{adx}f_{xbc} = 0 .
\end{align}
We want to arrange the amplitude for the NLSM in trivalent graphs as in \cref{eq:trivalent}, with the kinematic numerator satisfying the same identities as the curvature factor. Thus, the kinematic factor must satisfy the following identities
\begin{align}
	\label{eq:4ptConstraints1} 
	n(a,b,c,d) &= - n(b,a,c,d) , \\
	\label{eq:4ptConstraints2}
	n(a,b,c,d) &= - n(a,b,d,c) , \\
	\label{eq:4ptConstraints3}
	n(a,b,c,d) &= n(d,c,b,a) , \\
	\label{eq:4ptConstraints4}
	n(a,b,c,d) &= - n(a,c,d,b) - n(a,d,b,c) .
\end{align}
These constraints can be solved by using an ansatz for the kinematic numerator with unfixed parameters which---after imposing the constraints in \cref{eq:4ptConstraints1,eq:4ptConstraints2,eq:4ptConstraints3,eq:4ptConstraints4}---becomes
\begin{align}
	n(a,b,c,d) = - \frac{4}{3} s_{ab} (s_{ab} + 2 s_{bc}) .
\end{align}
Summing over all trivalent graphs, we match the NLSM amplitude
\begin{align}
	\mathcal{A}_{4} = \sum_{i \in \Gamma} \frac{ c_{i} n_{i} }{d_i} = R_{acbd} s_{ab} + R_{abcd} s_{ac} .
\end{align}
Now, if we replace the curvature factor with a second copy of the kinematic factor, we find the amplitude for the general (and special) Galileon theory
\begin{align}
	\mathcal{M}^{\rm Galileon}_{4} = \sum_{i \in \Gamma} \frac{n_i n_i}{d_i} \propto s_{ab} s_{ac} s_{bc} .
\end{align}
The NLSM satisfies the curvature-kinematics duality at four points and double copies to the Galileon theory. This agrees with the standard color-kinematics duality for the NLSM.

\subsection*{5-point amplitudes}

At four points we were able to reproduce the well-known result that the NLSM satisfies the color-kinematics duality and that its double copy is the special Galileon theory. We did this using curvatures rather than the adjoint color factors. 

At five points the situation is quite different. The NLSM amplitude for a symmetric coset vanishes (as do all odd-point amplitudes in that theory), while the odd-point amplitudes for nonsymmetric coset are nonzero. If one uses the same rules as for the adjoint color-kinematics duality, one can rule out any non-vanishing two-derivative scalar theory in the adjoint representation that satisfies the color-kinematics duality \cite{Bern:2019prr}. Therefore, as we will see below, some of the ingredients in the duality must be modified.

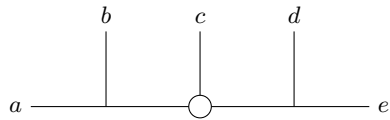
\begin{figure}
\centering
\begin{tikzpicture}
  \draw (-2.25,0) node[left] {$a$} -- (2.25,0) node[right] {$e$};
  \draw (-1.25,0) node[left] {} -- (-1.25,1) node[above] {$b$};
  \draw (1.25,0) node[above] {} -- (1.25,1) node[above] {$d$};
  \draw (0,0) node[above] {} -- (0,1) node[above] {$c$};
   \draw[fill=white] (0,0) circle (0.15) node[above=5pt]{};
\end{tikzpicture}
\caption{Trivalent graph for five particles.}
\label{fig:graph5pt}
\end{figure}

We start with the trivalent graph in \cref{fig:graph5pt}. The propagator is
\begin{align}
	d(a,b,c,d,e) = s_{ab} s_{de} .
\end{align}
One difference from before is that the graph has a new vertex. This is because we assign the following curvature to the graph
\begin{align}
	c(a,b,c,d,e) = \nabla_{c} R_{abde} .
\end{align}
The geometric structure as the following properties
\begin{align}
	\label{eq:5ptConstraints1}
	c(a,b,c,d,e) &= - c(a,b,c,e,d) , \\
	\label{eq:5ptConstraints2}
	c(a,b,c,d,e) &= - c(b,a,c,d,e) , \\
	\label{eq:5ptConstraints3}
	c(a,b,c,d,e) &= c(e,d,c,b,a) , \\
	\label{eq:5ptConstraints4}
	c(a,b,c,d,e) &= - c(a,d,c,e,b) - c(a,e,c,b,d) , \\
	\label{eq:5ptConstraints5}
	c(a,b,c,d,e) &= - c(b,c,a,d,e) - c(c,a,b,d,e) . 
\end{align}
In particular, the symmetry property in \cref{eq:5ptConstraints3} differs from the corresponding property for $f_{abx}f_{xcy}f_{yde}$ by a sign; $f_{abx}f_{xcy}f_{yde}=-f_{edy}f_{ycx}f_{xba}$. This symmetry property is what distinguishes the new vertex in \cref{fig:graph5pt} from the others.

Building color-kinematics dual amplitudes from color structures with different symmetry properties at 5 points was explored in ref.~\cite{Carrasco:2021ptp}. In one of the examples in that paper, the color structure was identified to be $f_{abx} d_{xcy} f_{yde}$, where $d_{abc} = {\rm Tr}[T_a \{T_b, T_c\}]$. For us, we see from \cref{eq:DR} that the curvature structure is coming from the antisymmetric combination of two types of structure constants, $f_{ija}$ and $f_{ijk}$. This combined with the antisymmetry of the middle structure constant results in an overall symmetric curvature under reflection. Note that the vertex is not fully symmetric; it is only symmetric when exchanging the two sides of the half-ladder diagram.

We build the kinematic numerators from an ansatz with 35 free parameters and demand that the kinematic numerator satisfies the same identities as the curvature structure;
\begin{align}
	n(a,b,c,d,e) &= - n(a,b,c,e,d) , \\
	n(a,b,c,d,e) &= - n(b,a,c,d,e) , \\
	n(a,b,c,d,e) &= n(e,d,c,b,a) , \\
	n(a,b,c,d,e) &= - n(a,d,c,e,b) - n(a,e,c,b,d) , \\
	n(a,b,c,d,e) &= - n(b,c,a,d,e) - n(c,a,b,d,e) . 
\end{align}
By imposing these constraints, we fix all but two of the parameters.

In ref.~\cite{Carrasco:2021ptp}, the kinematic numerator for the half-ladder graphs was seen to be proportional to a specific combination of momenta:
\begin{align}
	\label{eq:kinNumStructure}
	n(a,b,c,d,e) \propto (p_a - p_b)\cdot (p_d - p_e) .
\end{align}
Indeed, from our ansatz, we find a choice of the two free parameters that reproduces this structure. However, when we demand that the resulting amplitude should reproduce the NLSM amplitude, we must fix the free coefficients to a different linear combination. This means that our resulting kinematic numerator will not respect the structure in \cref{eq:kinNumStructure}. 

By summing all trivalent graphs, we reproduce the NLSM amplitude in \cref{eq:NLSM5pt},
\begin{align}
	\label{eq:5ptTrivalentSum}
	\mathcal{A}_{5} = \sum_{i\in\Gamma} \frac{c_{i} n_{i}}{d_{i}} .
\end{align}
The NLSM amplitude for nonsymmetric cosets respects the curvature-kinematics duality.

Next, we replace the curvature factor with a second copy of the kinematic factor. The result is
\begin{align}
	\mathcal{M}^{\rm Galileon}_{5} = \sum_{i\in\Gamma} \frac{n_i n_i}{d_i} \propto 
	\left( \sum_{a < b} s_{ab}^4 \right) - \frac{1}{4} \left( \sum_{a < b} s_{ab}^2\right)^2 .
\end{align}
The amplitude is from the general Galileon theory, not the special Galileon theory. 

This constructive approach can be extended to higher points. We have checked that the NLSM satisfies the curvature-kinematics duality through six points.

%%%%%%%%%%%%%%%%%%%%%%%%%%%%%%%%%%%%%%%%%%%%%%%%%%%%%%%
%%%%%%%%%%%%%%%% End Examples %%%%%%%%%%%%%%%%%%%%%%%%%
%%%%%%%%%%%%%%%%%%%%%%%%%%%%%%%%%%%%%%%%%%%%%%%%%%%%%%%

%%%%%%%%%%%%%%%%%%%%%%%%%%%%%%%%%%%%%%%%%%%%%%%%%%%%%%%
%%%%%%%%%%%%%%%% Begin Symmetric %%%%%%%%%%%%%%%%%%%%%%%%%
%%%%%%%%%%%%%%%%%%%%%%%%%%%%%%%%%%%%%%%%%%%%%%%%%%%%%%%

\section{An almost symmetric double copy}

The double copy comes in many guises. Instead of building the amplitude around adjoint color factors, a basis of symmetric color factors $d_{abc}$ was proposed in ref.~\cite{Carrasco:2022jxn}. These symmetric color factors satisfy fewer relations, which results in less cancellation between different terms in the sum over trivalent graphs.

We will explore the duality between symmetric curvatures and kinematics. At four points, the curvature factor for the trivalent graph is
\begin{align}
	c(a,b,c,d) &=  R_{cabd} + R_{cbad} .
\end{align}
The curvature is symmetric around each edge. Starting from an ansatz, we can find a dual kinematic factor
\begin{align}
	n(a,b,c,d) &=  (s_{ab})^2 .
\end{align}
By replacing the curvature with a second copy of the kinematic factor, we obtain the four-particle amplitude in the general (and special) Galileon amplitude.

Moving on to five particles, we can again start with a symmetric basis for the curvature terms. However, we must now address what symmetry property it should have under reflection. If we demand symmetry under reflection---as would be natural with the color factor $d_{abx} d_{xcy} d_{yde}$---we find a solution which does not double copy to the general Galileon theory. But if we instead demand antisymmetry under reflection, the color and kinematic factors which are dual to each other are
\begin{align}
	c(a,b,c,d,e) =& \nabla_{c} R_{adbe} - \nabla_{d} R_{acbe} - \nabla_{e} R_{adbc} , 
\end{align}
and
\begin{align}
	n(a,b,c,d,e) =& s_{ab} (s_{ac} + s_{bc} - s_{dc} - s_{ec}) s_{de} .
\end{align}
Plugging this back into the sum over trivalent graphs, \cref{eq:5ptTrivalentSum}, we recover the amplitude for the NLSM with a nonsymmetric coset. 
If we instead build an amplitude from two copies of the kinematic numerator, we again find that the double copy of the NLSM for nonsymmetric cosets is the general Galileon theory.

%%%%%%%%%%%%%%%%%%%%%%%%%%%%%%%%%%%%%%%%%%%%%%%%%%%%%%%
%%%%%%%%%%%%%%%% End Symmetric %%%%%%%%%%%%%%%%%%%%%%%%%
%%%%%%%%%%%%%%%%%%%%%%%%%%%%%%%%%%%%%%%%%%%%%%%%%%%%%%%

%%%%%%%%%%%%%%%%%%%%%%%%%%%%%%%%%%%%%%%%%%%%%%%%%%%%%%%
%%%%%%%%%%%%%%%% Begin Conclusion %%%%%%%%%%%%%%%%%%%%%
%%%%%%%%%%%%%%%%%%%%%%%%%%%%%%%%%%%%%%%%%%%%%%%%%%%%%%%

\section{Conclusion}

The double copy, which initially appeared to only concern very special theories, now implicates many quantum field theories. Gauge theories with various amounts of supersymmetry \cite{Chiodaroli:2014xia,Chiodaroli:2015rdg,Johansson:2017srf}, additional matter content \cite{Johansson:2015oia,Johansson:2019dnu,Carrasco:2020ywq}, large mass limit \cite{Haddad:2020tvs,Brandhuber:2021bsf}, and effective-field-theory extensions \cite{Chi:2021mio,Carrasco:2022lbm} all adhere to the tenets of the double copy. 

In this paper, we provide evidence for an extension of the class of color-kinematics dual theories. The double copy of the NLSM for nonsymmetric cosets is conjectured to be the general Galileon theory. The duality is built around the curvature of field space rather than the usual structure constants. As such, some of the algebraic properties differ. This is precisely what is needed to find a dual representation of the theory, which allows for a double copy. 

Many avenues are open for further exploration. First, and perhaps most pressing, is to lift the restriction of only having the curvature factors $R$ and $\nabla R$, and not higher curvature terms. For each $n$-point amplitude, there will be a new curvature term with its own algebraic properties. At the same time, the general Galileon has precisely one new higher-dimensional operator at each $n$-point with an unfixed coefficient. Thus, at each $n$-point there should be an independent double copy of the higher curvature terms which reproduces the higher-dimensional operators in the general Galileon theory. 

Another option is to search for a Lagrangian where the duality between the curvature and kinematics is manifest. This would amount to generalizing the Lagrangian in ref.~\cite{Cheung:2016prv} to nonsymmetric cosets. Achieving this would go a long way towards finding a proof of the conjectured duality. 

In the same vein, finding a bi-colored scalar theory which produces the graph structure for the geometric double copy would be very useful. This would necessarily extend the bi-adjoint scalar theory that underlies the usual adjoint color-kinematics duality. One would need two types of scalars to account for the interactions with the different structure constants $f_{aij}$ and $f_{ijk}$. 

Another interesting question to address is whether it is possible to embed the NLSM for nonsymmetric cosets into the scattering equation framework \cite{Cachazo:2013gna,Cachazo:2013hca,Cachazo:2013iea,Cachazo:2014nsa}. The color-ordering and number of independent ordered amplitudes is different from the symmetric coset case. This makes the extension of the scattering equation framework to nonsymmetric cosets highly nontrivial, if it is possible. We leave this for future study.

%%%%%%%%%%%%%%%%%%%%%%%%%%%%%%%%%%%%%%%%%%%%%%%%%%%%%%%
%%%%%%%%%%%%%%%% End Conclusion %%%%%%%%%%%%%%%%%%%%%%%
%%%%%%%%%%%%%%%%%%%%%%%%%%%%%%%%%%%%%%%%%%%%%%%%%%%%%%%

\acknowledgments{We thank Clifford Cheung and Julio Parra-Martinez for useful discussions. 
The Mathematica packages xAct \cite{xAct} and FiniteFieldSolve \cite{Mangan:2023eeb} where used during this work. }

\bibliography{bibliographyGeometryKinematics}

%apsrev4-2.bst 2019-01-14 (MD) hand-edited version of apsrev4-1.bst
%Control: key (0)
%Control: author (8) initials jnrlst
%Control: editor formatted (1) identically to author
%Control: production of article title (0) allowed
%Control: page (0) single
%Control: year (1) truncated
%Control: production of eprint (0) enabled
\begin{thebibliography}{34}%
\makeatletter
\providecommand \@ifxundefined [1]{%
 \@ifx{#1\undefined}
}%
\providecommand \@ifnum [1]{%
 \ifnum #1\expandafter \@firstoftwo
 \else \expandafter \@secondoftwo
 \fi
}%
\providecommand \@ifx [1]{%
 \ifx #1\expandafter \@firstoftwo
 \else \expandafter \@secondoftwo
 \fi
}%
\providecommand \natexlab [1]{#1}%
\providecommand \enquote  [1]{``#1''}%
\providecommand \bibnamefont  [1]{#1}%
\providecommand \bibfnamefont [1]{#1}%
\providecommand \citenamefont [1]{#1}%
\providecommand \href@noop [0]{\@secondoftwo}%
\providecommand \href [0]{\begingroup \@sanitize@url \@href}%
\providecommand \@href[1]{\@@startlink{#1}\@@href}%
\providecommand \@@href[1]{\endgroup#1\@@endlink}%
\providecommand \@sanitize@url [0]{\catcode `\\12\catcode `\$12\catcode
  `\&12\catcode `\#12\catcode `\^12\catcode `\_12\catcode `\%12\relax}%
\providecommand \@@startlink[1]{}%
\providecommand \@@endlink[0]{}%
\providecommand \url  [0]{\begingroup\@sanitize@url \@url }%
\providecommand \@url [1]{\endgroup\@href {#1}{\urlprefix }}%
\providecommand \urlprefix  [0]{URL }%
\providecommand \Eprint [0]{\href }%
\providecommand \doibase [0]{https://doi.org/}%
\providecommand \selectlanguage [0]{\@gobble}%
\providecommand \bibinfo  [0]{\@secondoftwo}%
\providecommand \bibfield  [0]{\@secondoftwo}%
\providecommand \translation [1]{[#1]}%
\providecommand \BibitemOpen [0]{}%
\providecommand \bibitemStop [0]{}%
\providecommand \bibitemNoStop [0]{.\EOS\space}%
\providecommand \EOS [0]{\spacefactor3000\relax}%
\providecommand \BibitemShut  [1]{\csname bibitem#1\endcsname}%
\let\auto@bib@innerbib\@empty
%</preamble>
\bibitem [{\citenamefont {Bern}\ \emph {et~al.}(2008)\citenamefont {Bern},
  \citenamefont {Carrasco},\ and\ \citenamefont {Johansson}}]{Bern:2008qj}%
  \BibitemOpen
  \bibfield  {author} {\bibinfo {author} {\bibfnamefont {Z.}~\bibnamefont
  {Bern}}, \bibinfo {author} {\bibfnamefont {J.~J.~M.}\ \bibnamefont
  {Carrasco}},\ and\ \bibinfo {author} {\bibfnamefont {H.}~\bibnamefont
  {Johansson}},\ }\bibfield  {title} {\bibinfo {title} {{New Relations for
  Gauge-Theory Amplitudes}},\ }\href
  {https://doi.org/10.1103/PhysRevD.78.085011} {\bibfield  {journal} {\bibinfo
  {journal} {Phys. Rev. D}\ }\textbf {\bibinfo {volume} {78}},\ \bibinfo
  {pages} {085011} (\bibinfo {year} {2008})},\ \Eprint
  {https://arxiv.org/abs/0805.3993} {arXiv:0805.3993 [hep-ph]} \BibitemShut
  {NoStop}%
\bibitem [{\citenamefont {Bern}\ \emph {et~al.}(2010)\citenamefont {Bern},
  \citenamefont {Carrasco},\ and\ \citenamefont {Johansson}}]{Bern:2010ue}%
  \BibitemOpen
  \bibfield  {author} {\bibinfo {author} {\bibfnamefont {Z.}~\bibnamefont
  {Bern}}, \bibinfo {author} {\bibfnamefont {J.~J.~M.}\ \bibnamefont
  {Carrasco}},\ and\ \bibinfo {author} {\bibfnamefont {H.}~\bibnamefont
  {Johansson}},\ }\bibfield  {title} {\bibinfo {title} {{Perturbative Quantum
  Gravity as a Double Copy of Gauge Theory}},\ }\href
  {https://doi.org/10.1103/PhysRevLett.105.061602} {\bibfield  {journal}
  {\bibinfo  {journal} {Phys. Rev. Lett.}\ }\textbf {\bibinfo {volume} {105}},\
  \bibinfo {pages} {061602} (\bibinfo {year} {2010})},\ \Eprint
  {https://arxiv.org/abs/1004.0476} {arXiv:1004.0476 [hep-th]} \BibitemShut
  {NoStop}%
\bibitem [{\citenamefont {Kawai}\ \emph {et~al.}(1986)\citenamefont {Kawai},
  \citenamefont {Lewellen},\ and\ \citenamefont {Tye}}]{Kawai:1985xq}%
  \BibitemOpen
  \bibfield  {author} {\bibinfo {author} {\bibfnamefont {H.}~\bibnamefont
  {Kawai}}, \bibinfo {author} {\bibfnamefont {D.~C.}\ \bibnamefont
  {Lewellen}},\ and\ \bibinfo {author} {\bibfnamefont {S.~H.~H.}\ \bibnamefont
  {Tye}},\ }\bibfield  {title} {\bibinfo {title} {{A Relation Between Tree
  Amplitudes of Closed and Open Strings}},\ }\href
  {https://doi.org/10.1016/0550-3213(86)90362-7} {\bibfield  {journal}
  {\bibinfo  {journal} {Nucl. Phys. B}\ }\textbf {\bibinfo {volume} {269}},\
  \bibinfo {pages} {1} (\bibinfo {year} {1986})}\BibitemShut {NoStop}%
\bibitem [{\citenamefont {Bern}\ \emph {et~al.}(2019)\citenamefont {Bern},
  \citenamefont {Carrasco}, \citenamefont {Chiodaroli}, \citenamefont
  {Johansson},\ and\ \citenamefont {Roiban}}]{Bern:2019prr}%
  \BibitemOpen
  \bibfield  {author} {\bibinfo {author} {\bibfnamefont {Z.}~\bibnamefont
  {Bern}}, \bibinfo {author} {\bibfnamefont {J.~J.}\ \bibnamefont {Carrasco}},
  \bibinfo {author} {\bibfnamefont {M.}~\bibnamefont {Chiodaroli}}, \bibinfo
  {author} {\bibfnamefont {H.}~\bibnamefont {Johansson}},\ and\ \bibinfo
  {author} {\bibfnamefont {R.}~\bibnamefont {Roiban}},\ }\bibfield  {title}
  {\bibinfo {title} {{The Duality Between Color and Kinematics and its
  Applications}},\ }\href@noop {} {\  (\bibinfo {year} {2019})},\ \Eprint
  {https://arxiv.org/abs/1909.01358} {arXiv:1909.01358 [hep-th]} \BibitemShut
  {NoStop}%
\bibitem [{\citenamefont {Bern}\ \emph {et~al.}(2022)\citenamefont {Bern},
  \citenamefont {Carrasco}, \citenamefont {Chiodaroli}, \citenamefont
  {Johansson},\ and\ \citenamefont {Roiban}}]{Bern:2022wqg}%
  \BibitemOpen
  \bibfield  {author} {\bibinfo {author} {\bibfnamefont {Z.}~\bibnamefont
  {Bern}}, \bibinfo {author} {\bibfnamefont {J.~J.}\ \bibnamefont {Carrasco}},
  \bibinfo {author} {\bibfnamefont {M.}~\bibnamefont {Chiodaroli}}, \bibinfo
  {author} {\bibfnamefont {H.}~\bibnamefont {Johansson}},\ and\ \bibinfo
  {author} {\bibfnamefont {R.}~\bibnamefont {Roiban}},\ }\bibfield  {title}
  {\bibinfo {title} {{The SAGEX review on scattering amplitudes Chapter 2: An
  invitation to color-kinematics duality and the double copy}},\ }\href
  {https://doi.org/10.1088/1751-8121/ac93cf} {\bibfield  {journal} {\bibinfo
  {journal} {J. Phys. A}\ }\textbf {\bibinfo {volume} {55}},\ \bibinfo {pages}
  {443003} (\bibinfo {year} {2022})},\ \Eprint
  {https://arxiv.org/abs/2203.13013} {arXiv:2203.13013 [hep-th]} \BibitemShut
  {NoStop}%
\bibitem [{\citenamefont {Adler}(1965)}]{Adler:1964um}%
  \BibitemOpen
  \bibfield  {author} {\bibinfo {author} {\bibfnamefont {S.~L.}\ \bibnamefont
  {Adler}},\ }\bibfield  {title} {\bibinfo {title} {{Consistency conditions on
  the strong interactions implied by a partially conserved axial vector
  current}},\ }\href {https://doi.org/10.1103/PhysRev.137.B1022} {\bibfield
  {journal} {\bibinfo  {journal} {Phys. Rev.}\ }\textbf {\bibinfo {volume}
  {137}},\ \bibinfo {pages} {B1022} (\bibinfo {year} {1965})}\BibitemShut
  {NoStop}%
\bibitem [{\citenamefont {Cheung}\ \emph {et~al.}(2015)\citenamefont {Cheung},
  \citenamefont {Kampf}, \citenamefont {Novotny},\ and\ \citenamefont
  {Trnka}}]{Cheung:2014dqa}%
  \BibitemOpen
  \bibfield  {author} {\bibinfo {author} {\bibfnamefont {C.}~\bibnamefont
  {Cheung}}, \bibinfo {author} {\bibfnamefont {K.}~\bibnamefont {Kampf}},
  \bibinfo {author} {\bibfnamefont {J.}~\bibnamefont {Novotny}},\ and\ \bibinfo
  {author} {\bibfnamefont {J.}~\bibnamefont {Trnka}},\ }\bibfield  {title}
  {\bibinfo {title} {{Effective Field Theories from Soft Limits of Scattering
  Amplitudes}},\ }\href {https://doi.org/10.1103/PhysRevLett.114.221602}
  {\bibfield  {journal} {\bibinfo  {journal} {Phys. Rev. Lett.}\ }\textbf
  {\bibinfo {volume} {114}},\ \bibinfo {pages} {221602} (\bibinfo {year}
  {2015})},\ \Eprint {https://arxiv.org/abs/1412.4095} {arXiv:1412.4095
  [hep-th]} \BibitemShut {NoStop}%
\bibitem [{\citenamefont {Hinterbichler}\ and\ \citenamefont
  {Joyce}(2015)}]{Hinterbichler:2015pqa}%
  \BibitemOpen
  \bibfield  {author} {\bibinfo {author} {\bibfnamefont {K.}~\bibnamefont
  {Hinterbichler}}\ and\ \bibinfo {author} {\bibfnamefont {A.}~\bibnamefont
  {Joyce}},\ }\bibfield  {title} {\bibinfo {title} {{Hidden symmetry of the
  Galileon}},\ }\href {https://doi.org/10.1103/PhysRevD.92.023503} {\bibfield
  {journal} {\bibinfo  {journal} {Phys. Rev. D}\ }\textbf {\bibinfo {volume}
  {92}},\ \bibinfo {pages} {023503} (\bibinfo {year} {2015})},\ \Eprint
  {https://arxiv.org/abs/1501.07600} {arXiv:1501.07600 [hep-th]} \BibitemShut
  {NoStop}%
\bibitem [{\citenamefont {Cheung}\ \emph {et~al.}(2017)\citenamefont {Cheung},
  \citenamefont {Kampf}, \citenamefont {Novotny}, \citenamefont {Shen},\ and\
  \citenamefont {Trnka}}]{Cheung:2016drk}%
  \BibitemOpen
  \bibfield  {author} {\bibinfo {author} {\bibfnamefont {C.}~\bibnamefont
  {Cheung}}, \bibinfo {author} {\bibfnamefont {K.}~\bibnamefont {Kampf}},
  \bibinfo {author} {\bibfnamefont {J.}~\bibnamefont {Novotny}}, \bibinfo
  {author} {\bibfnamefont {C.-H.}\ \bibnamefont {Shen}},\ and\ \bibinfo
  {author} {\bibfnamefont {J.}~\bibnamefont {Trnka}},\ }\bibfield  {title}
  {\bibinfo {title} {{A Periodic Table of Effective Field Theories}},\ }\href
  {https://doi.org/10.1007/JHEP02(2017)020} {\bibfield  {journal} {\bibinfo
  {journal} {JHEP}\ }\textbf {\bibinfo {volume} {02}},\ \bibinfo {pages}
  {020}},\ \Eprint {https://arxiv.org/abs/1611.03137} {arXiv:1611.03137
  [hep-th]} \BibitemShut {NoStop}%
\bibitem [{\citenamefont {Cheung}\ \emph
  {et~al.}(2022{\natexlab{a}})\citenamefont {Cheung}, \citenamefont {Helset},\
  and\ \citenamefont {Parra-Martinez}}]{Cheung:2021yog}%
  \BibitemOpen
  \bibfield  {author} {\bibinfo {author} {\bibfnamefont {C.}~\bibnamefont
  {Cheung}}, \bibinfo {author} {\bibfnamefont {A.}~\bibnamefont {Helset}},\
  and\ \bibinfo {author} {\bibfnamefont {J.}~\bibnamefont {Parra-Martinez}},\
  }\bibfield  {title} {\bibinfo {title} {{Geometric soft theorems}},\ }\href
  {https://doi.org/10.1007/JHEP04(2022)011} {\bibfield  {journal} {\bibinfo
  {journal} {JHEP}\ }\textbf {\bibinfo {volume} {04}},\ \bibinfo {pages}
  {011}},\ \Eprint {https://arxiv.org/abs/2111.03045} {arXiv:2111.03045
  [hep-th]} \BibitemShut {NoStop}%
\bibitem [{\citenamefont {Derda}\ \emph {et~al.}(2024)\citenamefont {Derda},
  \citenamefont {Helset},\ and\ \citenamefont
  {Parra-Martinez}}]{Derda:2024jvo}%
  \BibitemOpen
  \bibfield  {author} {\bibinfo {author} {\bibfnamefont {M.}~\bibnamefont
  {Derda}}, \bibinfo {author} {\bibfnamefont {A.}~\bibnamefont {Helset}},\ and\
  \bibinfo {author} {\bibfnamefont {J.}~\bibnamefont {Parra-Martinez}},\
  }\bibfield  {title} {\bibinfo {title} {{Soft Scalars in Effective Field
  Theory}},\ }\href@noop {} {\  (\bibinfo {year} {2024})},\ \Eprint
  {https://arxiv.org/abs/2403.12142} {arXiv:2403.12142 [hep-th]} \BibitemShut
  {NoStop}%
\bibitem [{\citenamefont {Coleman}\ \emph {et~al.}(1969)\citenamefont
  {Coleman}, \citenamefont {Wess},\ and\ \citenamefont
  {Zumino}}]{Coleman:1969sm}%
  \BibitemOpen
  \bibfield  {author} {\bibinfo {author} {\bibfnamefont {S.~R.}\ \bibnamefont
  {Coleman}}, \bibinfo {author} {\bibfnamefont {J.}~\bibnamefont {Wess}},\ and\
  \bibinfo {author} {\bibfnamefont {B.}~\bibnamefont {Zumino}},\ }\bibfield
  {title} {\bibinfo {title} {{Structure of phenomenological Lagrangians. 1.}},\
  }\href {https://doi.org/10.1103/PhysRev.177.2239} {\bibfield  {journal}
  {\bibinfo  {journal} {Phys. Rev.}\ }\textbf {\bibinfo {volume} {177}},\
  \bibinfo {pages} {2239} (\bibinfo {year} {1969})}\BibitemShut {NoStop}%
\bibitem [{\citenamefont {Callan}\ \emph {et~al.}(1969)\citenamefont {Callan},
  \citenamefont {Coleman}, \citenamefont {Wess},\ and\ \citenamefont
  {Zumino}}]{Callan:1969sn}%
  \BibitemOpen
  \bibfield  {author} {\bibinfo {author} {\bibfnamefont {C.~G.}\ \bibnamefont
  {Callan}, \bibfnamefont {Jr.}}, \bibinfo {author} {\bibfnamefont {S.~R.}\
  \bibnamefont {Coleman}}, \bibinfo {author} {\bibfnamefont {J.}~\bibnamefont
  {Wess}},\ and\ \bibinfo {author} {\bibfnamefont {B.}~\bibnamefont {Zumino}},\
  }\bibfield  {title} {\bibinfo {title} {{Structure of phenomenological
  Lagrangians. 2.}},\ }\href {https://doi.org/10.1103/PhysRev.177.2247}
  {\bibfield  {journal} {\bibinfo  {journal} {Phys. Rev.}\ }\textbf {\bibinfo
  {volume} {177}},\ \bibinfo {pages} {2247} (\bibinfo {year}
  {1969})}\BibitemShut {NoStop}%
\bibitem [{\citenamefont {Alonso}\ \emph {et~al.}(2016)\citenamefont {Alonso},
  \citenamefont {Jenkins},\ and\ \citenamefont {Manohar}}]{Alonso:2016oah}%
  \BibitemOpen
  \bibfield  {author} {\bibinfo {author} {\bibfnamefont {R.}~\bibnamefont
  {Alonso}}, \bibinfo {author} {\bibfnamefont {E.~E.}\ \bibnamefont
  {Jenkins}},\ and\ \bibinfo {author} {\bibfnamefont {A.~V.}\ \bibnamefont
  {Manohar}},\ }\bibfield  {title} {\bibinfo {title} {{Geometry of the Scalar
  Sector}},\ }\href {https://doi.org/10.1007/JHEP08(2016)101} {\bibfield
  {journal} {\bibinfo  {journal} {JHEP}\ }\textbf {\bibinfo {volume} {08}},\
  \bibinfo {pages} {101}},\ \Eprint {https://arxiv.org/abs/1605.03602}
  {arXiv:1605.03602 [hep-ph]} \BibitemShut {NoStop}%
\bibitem [{\citenamefont {Cheung}\ \emph
  {et~al.}(2022{\natexlab{b}})\citenamefont {Cheung}, \citenamefont {Helset},\
  and\ \citenamefont {Parra-Martinez}}]{Cheung:2022vnd}%
  \BibitemOpen
  \bibfield  {author} {\bibinfo {author} {\bibfnamefont {C.}~\bibnamefont
  {Cheung}}, \bibinfo {author} {\bibfnamefont {A.}~\bibnamefont {Helset}},\
  and\ \bibinfo {author} {\bibfnamefont {J.}~\bibnamefont {Parra-Martinez}},\
  }\bibfield  {title} {\bibinfo {title} {{Geometry-kinematics duality}},\
  }\href {https://doi.org/10.1103/PhysRevD.106.045016} {\bibfield  {journal}
  {\bibinfo  {journal} {Phys. Rev. D}\ }\textbf {\bibinfo {volume} {106}},\
  \bibinfo {pages} {045016} (\bibinfo {year} {2022}{\natexlab{b}})},\ \Eprint
  {https://arxiv.org/abs/2202.06972} {arXiv:2202.06972 [hep-th]} \BibitemShut
  {NoStop}%
\bibitem [{\citenamefont {Carrasco}\ \emph {et~al.}(2021)\citenamefont
  {Carrasco}, \citenamefont {Rodina},\ and\ \citenamefont
  {Zekioglu}}]{Carrasco:2021ptp}%
  \BibitemOpen
  \bibfield  {author} {\bibinfo {author} {\bibfnamefont {J.~J.~M.}\
  \bibnamefont {Carrasco}}, \bibinfo {author} {\bibfnamefont {L.}~\bibnamefont
  {Rodina}},\ and\ \bibinfo {author} {\bibfnamefont {S.}~\bibnamefont
  {Zekioglu}},\ }\bibfield  {title} {\bibinfo {title} {{Composing effective
  prediction at five points}},\ }\href
  {https://doi.org/10.1007/JHEP06(2021)169} {\bibfield  {journal} {\bibinfo
  {journal} {JHEP}\ }\textbf {\bibinfo {volume} {06}},\ \bibinfo {pages}
  {169}},\ \Eprint {https://arxiv.org/abs/2104.08370} {arXiv:2104.08370
  [hep-th]} \BibitemShut {NoStop}%
\bibitem [{\citenamefont {Carrasco}\ and\ \citenamefont
  {Pavao}(2023)}]{Carrasco:2022jxn}%
  \BibitemOpen
  \bibfield  {author} {\bibinfo {author} {\bibfnamefont {J.~J.~M.}\
  \bibnamefont {Carrasco}}\ and\ \bibinfo {author} {\bibfnamefont {N.~H.}\
  \bibnamefont {Pavao}},\ }\bibfield  {title} {\bibinfo {title} {{Virtues of a
  symmetric-structure double copy}},\ }\href
  {https://doi.org/10.1103/PhysRevD.107.065005} {\bibfield  {journal} {\bibinfo
   {journal} {Phys. Rev. D}\ }\textbf {\bibinfo {volume} {107}},\ \bibinfo
  {pages} {065005} (\bibinfo {year} {2023})},\ \Eprint
  {https://arxiv.org/abs/2211.04431} {arXiv:2211.04431 [hep-th]} \BibitemShut
  {NoStop}%
\bibitem [{\citenamefont {Chiodaroli}\ \emph {et~al.}(2015)\citenamefont
  {Chiodaroli}, \citenamefont {G\"unaydin}, \citenamefont {Johansson},\ and\
  \citenamefont {Roiban}}]{Chiodaroli:2014xia}%
  \BibitemOpen
  \bibfield  {author} {\bibinfo {author} {\bibfnamefont {M.}~\bibnamefont
  {Chiodaroli}}, \bibinfo {author} {\bibfnamefont {M.}~\bibnamefont
  {G\"unaydin}}, \bibinfo {author} {\bibfnamefont {H.}~\bibnamefont
  {Johansson}},\ and\ \bibinfo {author} {\bibfnamefont {R.}~\bibnamefont
  {Roiban}},\ }\bibfield  {title} {\bibinfo {title} {{Scattering amplitudes in
  $ \mathcal{N}=2 $ Maxwell-Einstein and Yang-Mills/Einstein supergravity}},\
  }\href {https://doi.org/10.1007/JHEP01(2015)081} {\bibfield  {journal}
  {\bibinfo  {journal} {JHEP}\ }\textbf {\bibinfo {volume} {01}},\ \bibinfo
  {pages} {081}},\ \Eprint {https://arxiv.org/abs/1408.0764} {arXiv:1408.0764
  [hep-th]} \BibitemShut {NoStop}%
\bibitem [{\citenamefont {Chiodaroli}\ \emph {et~al.}(2017)\citenamefont
  {Chiodaroli}, \citenamefont {Gunaydin}, \citenamefont {Johansson},\ and\
  \citenamefont {Roiban}}]{Chiodaroli:2015rdg}%
  \BibitemOpen
  \bibfield  {author} {\bibinfo {author} {\bibfnamefont {M.}~\bibnamefont
  {Chiodaroli}}, \bibinfo {author} {\bibfnamefont {M.}~\bibnamefont
  {Gunaydin}}, \bibinfo {author} {\bibfnamefont {H.}~\bibnamefont
  {Johansson}},\ and\ \bibinfo {author} {\bibfnamefont {R.}~\bibnamefont
  {Roiban}},\ }\bibfield  {title} {\bibinfo {title} {{Spontaneously Broken
  Yang-Mills-Einstein Supergravities as Double Copies}},\ }\href
  {https://doi.org/10.1007/JHEP06(2017)064} {\bibfield  {journal} {\bibinfo
  {journal} {JHEP}\ }\textbf {\bibinfo {volume} {06}},\ \bibinfo {pages}
  {064}},\ \Eprint {https://arxiv.org/abs/1511.01740} {arXiv:1511.01740
  [hep-th]} \BibitemShut {NoStop}%
\bibitem [{\citenamefont {Johansson}\ and\ \citenamefont
  {Nohle}(2017)}]{Johansson:2017srf}%
  \BibitemOpen
  \bibfield  {author} {\bibinfo {author} {\bibfnamefont {H.}~\bibnamefont
  {Johansson}}\ and\ \bibinfo {author} {\bibfnamefont {J.}~\bibnamefont
  {Nohle}},\ }\bibfield  {title} {\bibinfo {title} {{Conformal Gravity from
  Gauge Theory}},\ }\href@noop {} {\  (\bibinfo {year} {2017})},\ \Eprint
  {https://arxiv.org/abs/1707.02965} {arXiv:1707.02965 [hep-th]} \BibitemShut
  {NoStop}%
\bibitem [{\citenamefont {Johansson}\ and\ \citenamefont
  {Ochirov}(2016)}]{Johansson:2015oia}%
  \BibitemOpen
  \bibfield  {author} {\bibinfo {author} {\bibfnamefont {H.}~\bibnamefont
  {Johansson}}\ and\ \bibinfo {author} {\bibfnamefont {A.}~\bibnamefont
  {Ochirov}},\ }\bibfield  {title} {\bibinfo {title} {{Color-Kinematics Duality
  for QCD Amplitudes}},\ }\href {https://doi.org/10.1007/JHEP01(2016)170}
  {\bibfield  {journal} {\bibinfo  {journal} {JHEP}\ }\textbf {\bibinfo
  {volume} {01}},\ \bibinfo {pages} {170}},\ \Eprint
  {https://arxiv.org/abs/1507.00332} {arXiv:1507.00332 [hep-ph]} \BibitemShut
  {NoStop}%
\bibitem [{\citenamefont {Johansson}\ and\ \citenamefont
  {Ochirov}(2019)}]{Johansson:2019dnu}%
  \BibitemOpen
  \bibfield  {author} {\bibinfo {author} {\bibfnamefont {H.}~\bibnamefont
  {Johansson}}\ and\ \bibinfo {author} {\bibfnamefont {A.}~\bibnamefont
  {Ochirov}},\ }\bibfield  {title} {\bibinfo {title} {{Double copy for massive
  quantum particles with spin}},\ }\href
  {https://doi.org/10.1007/JHEP09(2019)040} {\bibfield  {journal} {\bibinfo
  {journal} {JHEP}\ }\textbf {\bibinfo {volume} {09}},\ \bibinfo {pages}
  {040}},\ \Eprint {https://arxiv.org/abs/1906.12292} {arXiv:1906.12292
  [hep-th]} \BibitemShut {NoStop}%
\bibitem [{\citenamefont {Carrasco}\ and\ \citenamefont
  {Vazquez-Holm}(2021)}]{Carrasco:2020ywq}%
  \BibitemOpen
  \bibfield  {author} {\bibinfo {author} {\bibfnamefont {J.~J.~M.}\
  \bibnamefont {Carrasco}}\ and\ \bibinfo {author} {\bibfnamefont {I.~A.}\
  \bibnamefont {Vazquez-Holm}},\ }\bibfield  {title} {\bibinfo {title}
  {{Loop-Level Double-Copy for Massive Quantum Particles}},\ }\href
  {https://doi.org/10.1103/PhysRevD.103.045002} {\bibfield  {journal} {\bibinfo
   {journal} {Phys. Rev. D}\ }\textbf {\bibinfo {volume} {103}},\ \bibinfo
  {pages} {045002} (\bibinfo {year} {2021})},\ \Eprint
  {https://arxiv.org/abs/2010.13435} {arXiv:2010.13435 [hep-th]} \BibitemShut
  {NoStop}%
\bibitem [{\citenamefont {Haddad}\ and\ \citenamefont
  {Helset}(2020)}]{Haddad:2020tvs}%
  \BibitemOpen
  \bibfield  {author} {\bibinfo {author} {\bibfnamefont {K.}~\bibnamefont
  {Haddad}}\ and\ \bibinfo {author} {\bibfnamefont {A.}~\bibnamefont
  {Helset}},\ }\bibfield  {title} {\bibinfo {title} {{The double copy for heavy
  particles}},\ }\href {https://doi.org/10.1103/PhysRevLett.125.181603}
  {\bibfield  {journal} {\bibinfo  {journal} {Phys. Rev. Lett.}\ }\textbf
  {\bibinfo {volume} {125}},\ \bibinfo {pages} {181603} (\bibinfo {year}
  {2020})},\ \Eprint {https://arxiv.org/abs/2005.13897} {arXiv:2005.13897
  [hep-th]} \BibitemShut {NoStop}%
\bibitem [{\citenamefont {Brandhuber}\ \emph {et~al.}(2022)\citenamefont
  {Brandhuber}, \citenamefont {Chen}, \citenamefont {Johansson}, \citenamefont
  {Travaglini},\ and\ \citenamefont {Wen}}]{Brandhuber:2021bsf}%
  \BibitemOpen
  \bibfield  {author} {\bibinfo {author} {\bibfnamefont {A.}~\bibnamefont
  {Brandhuber}}, \bibinfo {author} {\bibfnamefont {G.}~\bibnamefont {Chen}},
  \bibinfo {author} {\bibfnamefont {H.}~\bibnamefont {Johansson}}, \bibinfo
  {author} {\bibfnamefont {G.}~\bibnamefont {Travaglini}},\ and\ \bibinfo
  {author} {\bibfnamefont {C.}~\bibnamefont {Wen}},\ }\bibfield  {title}
  {\bibinfo {title} {{Kinematic Hopf Algebra for Bern-Carrasco-Johansson
  Numerators in Heavy-Mass Effective Field Theory and Yang-Mills Theory}},\
  }\href {https://doi.org/10.1103/PhysRevLett.128.121601} {\bibfield  {journal}
  {\bibinfo  {journal} {Phys. Rev. Lett.}\ }\textbf {\bibinfo {volume} {128}},\
  \bibinfo {pages} {121601} (\bibinfo {year} {2022})},\ \Eprint
  {https://arxiv.org/abs/2111.15649} {arXiv:2111.15649 [hep-th]} \BibitemShut
  {NoStop}%
\bibitem [{\citenamefont {Chi}\ \emph {et~al.}(2022)\citenamefont {Chi},
  \citenamefont {Elvang}, \citenamefont {Herderschee}, \citenamefont {Jones},\
  and\ \citenamefont {Paranjape}}]{Chi:2021mio}%
  \BibitemOpen
  \bibfield  {author} {\bibinfo {author} {\bibfnamefont {H.-H.}\ \bibnamefont
  {Chi}}, \bibinfo {author} {\bibfnamefont {H.}~\bibnamefont {Elvang}},
  \bibinfo {author} {\bibfnamefont {A.}~\bibnamefont {Herderschee}}, \bibinfo
  {author} {\bibfnamefont {C.~R.~T.}\ \bibnamefont {Jones}},\ and\ \bibinfo
  {author} {\bibfnamefont {S.}~\bibnamefont {Paranjape}},\ }\bibfield  {title}
  {\bibinfo {title} {{Generalizations of the double-copy: the KLT bootstrap}},\
  }\href {https://doi.org/10.1007/JHEP03(2022)077} {\bibfield  {journal}
  {\bibinfo  {journal} {JHEP}\ }\textbf {\bibinfo {volume} {03}},\ \bibinfo
  {pages} {077}},\ \Eprint {https://arxiv.org/abs/2106.12600} {arXiv:2106.12600
  [hep-th]} \BibitemShut {NoStop}%
\bibitem [{\citenamefont {Carrasco}\ \emph {et~al.}(2023)\citenamefont
  {Carrasco}, \citenamefont {Lewandowski},\ and\ \citenamefont
  {Pavao}}]{Carrasco:2022lbm}%
  \BibitemOpen
  \bibfield  {author} {\bibinfo {author} {\bibfnamefont {J.~J.~M.}\
  \bibnamefont {Carrasco}}, \bibinfo {author} {\bibfnamefont {M.}~\bibnamefont
  {Lewandowski}},\ and\ \bibinfo {author} {\bibfnamefont {N.~H.}\ \bibnamefont
  {Pavao}},\ }\bibfield  {title} {\bibinfo {title} {{Color-Dual Fates of F3,
  R3, and N=4 Supergravity}},\ }\href
  {https://doi.org/10.1103/PhysRevLett.131.051601} {\bibfield  {journal}
  {\bibinfo  {journal} {Phys. Rev. Lett.}\ }\textbf {\bibinfo {volume} {131}},\
  \bibinfo {pages} {051601} (\bibinfo {year} {2023})},\ \Eprint
  {https://arxiv.org/abs/2203.03592} {arXiv:2203.03592 [hep-th]} \BibitemShut
  {NoStop}%
\bibitem [{\citenamefont {Cheung}\ and\ \citenamefont
  {Shen}(2017)}]{Cheung:2016prv}%
  \BibitemOpen
  \bibfield  {author} {\bibinfo {author} {\bibfnamefont {C.}~\bibnamefont
  {Cheung}}\ and\ \bibinfo {author} {\bibfnamefont {C.-H.}\ \bibnamefont
  {Shen}},\ }\bibfield  {title} {\bibinfo {title} {{Symmetry for
  Flavor-Kinematics Duality from an Action}},\ }\href
  {https://doi.org/10.1103/PhysRevLett.118.121601} {\bibfield  {journal}
  {\bibinfo  {journal} {Phys. Rev. Lett.}\ }\textbf {\bibinfo {volume} {118}},\
  \bibinfo {pages} {121601} (\bibinfo {year} {2017})},\ \Eprint
  {https://arxiv.org/abs/1612.00868} {arXiv:1612.00868 [hep-th]} \BibitemShut
  {NoStop}%
\bibitem [{\citenamefont {Cachazo}\ \emph
  {et~al.}(2014{\natexlab{a}})\citenamefont {Cachazo}, \citenamefont {He},\
  and\ \citenamefont {Yuan}}]{Cachazo:2013gna}%
  \BibitemOpen
  \bibfield  {author} {\bibinfo {author} {\bibfnamefont {F.}~\bibnamefont
  {Cachazo}}, \bibinfo {author} {\bibfnamefont {S.}~\bibnamefont {He}},\ and\
  \bibinfo {author} {\bibfnamefont {E.~Y.}\ \bibnamefont {Yuan}},\ }\bibfield
  {title} {\bibinfo {title} {{Scattering equations and Kawai-Lewellen-Tye
  orthogonality}},\ }\href {https://doi.org/10.1103/PhysRevD.90.065001}
  {\bibfield  {journal} {\bibinfo  {journal} {Phys. Rev. D}\ }\textbf {\bibinfo
  {volume} {90}},\ \bibinfo {pages} {065001} (\bibinfo {year}
  {2014}{\natexlab{a}})},\ \Eprint {https://arxiv.org/abs/1306.6575}
  {arXiv:1306.6575 [hep-th]} \BibitemShut {NoStop}%
\bibitem [{\citenamefont {Cachazo}\ \emph
  {et~al.}(2014{\natexlab{b}})\citenamefont {Cachazo}, \citenamefont {He},\
  and\ \citenamefont {Yuan}}]{Cachazo:2013hca}%
  \BibitemOpen
  \bibfield  {author} {\bibinfo {author} {\bibfnamefont {F.}~\bibnamefont
  {Cachazo}}, \bibinfo {author} {\bibfnamefont {S.}~\bibnamefont {He}},\ and\
  \bibinfo {author} {\bibfnamefont {E.~Y.}\ \bibnamefont {Yuan}},\ }\bibfield
  {title} {\bibinfo {title} {{Scattering of Massless Particles in Arbitrary
  Dimensions}},\ }\href {https://doi.org/10.1103/PhysRevLett.113.171601}
  {\bibfield  {journal} {\bibinfo  {journal} {Phys. Rev. Lett.}\ }\textbf
  {\bibinfo {volume} {113}},\ \bibinfo {pages} {171601} (\bibinfo {year}
  {2014}{\natexlab{b}})},\ \Eprint {https://arxiv.org/abs/1307.2199}
  {arXiv:1307.2199 [hep-th]} \BibitemShut {NoStop}%
\bibitem [{\citenamefont {Cachazo}\ \emph
  {et~al.}(2014{\natexlab{c}})\citenamefont {Cachazo}, \citenamefont {He},\
  and\ \citenamefont {Yuan}}]{Cachazo:2013iea}%
  \BibitemOpen
  \bibfield  {author} {\bibinfo {author} {\bibfnamefont {F.}~\bibnamefont
  {Cachazo}}, \bibinfo {author} {\bibfnamefont {S.}~\bibnamefont {He}},\ and\
  \bibinfo {author} {\bibfnamefont {E.~Y.}\ \bibnamefont {Yuan}},\ }\bibfield
  {title} {\bibinfo {title} {{Scattering of Massless Particles: Scalars, Gluons
  and Gravitons}},\ }\href {https://doi.org/10.1007/JHEP07(2014)033} {\bibfield
   {journal} {\bibinfo  {journal} {JHEP}\ }\textbf {\bibinfo {volume} {07}},\
  \bibinfo {pages} {033}},\ \Eprint {https://arxiv.org/abs/1309.0885}
  {arXiv:1309.0885 [hep-th]} \BibitemShut {NoStop}%
\bibitem [{\citenamefont {Cachazo}\ \emph {et~al.}(2015)\citenamefont
  {Cachazo}, \citenamefont {He},\ and\ \citenamefont {Yuan}}]{Cachazo:2014nsa}%
  \BibitemOpen
  \bibfield  {author} {\bibinfo {author} {\bibfnamefont {F.}~\bibnamefont
  {Cachazo}}, \bibinfo {author} {\bibfnamefont {S.}~\bibnamefont {He}},\ and\
  \bibinfo {author} {\bibfnamefont {E.~Y.}\ \bibnamefont {Yuan}},\ }\bibfield
  {title} {\bibinfo {title} {{Einstein-Yang-Mills Scattering Amplitudes From
  Scattering Equations}},\ }\href {https://doi.org/10.1007/JHEP01(2015)121}
  {\bibfield  {journal} {\bibinfo  {journal} {JHEP}\ }\textbf {\bibinfo
  {volume} {01}},\ \bibinfo {pages} {121}},\ \Eprint
  {https://arxiv.org/abs/1409.8256} {arXiv:1409.8256 [hep-th]} \BibitemShut
  {NoStop}%
\bibitem [{\citenamefont {Mart\'in-Garc\'ia}()}]{xAct}%
  \BibitemOpen
  \bibfield  {author} {\bibinfo {author} {\bibfnamefont {J.~M.}\ \bibnamefont
  {Mart\'in-Garc\'ia}},\ }\bibfield  {title} {\bibinfo {title} {{xAct:
  Efficient tensor computer algebra for the Wolfram Language}},\ }\href
  {http://www.xact.es/index.html} {\bibinfo  {journal} {www.xact.es}\
  }\BibitemShut {NoStop}%
\bibitem [{\citenamefont {Mangan}(2024)}]{Mangan:2023eeb}%
  \BibitemOpen
\bibfield  {journal} {  }\bibfield  {author} {\bibinfo {author} {\bibfnamefont
  {J.}~\bibnamefont {Mangan}},\ }\bibfield  {title} {\bibinfo {title}
  {{FiniteFieldSolve: Exactly solving large linear systems in high-energy
  theory}},\ }\href {https://doi.org/10.1016/j.cpc.2024.109171} {\bibfield
  {journal} {\bibinfo  {journal} {Comput. Phys. Commun.}\ }\textbf {\bibinfo
  {volume} {300}},\ \bibinfo {pages} {109171} (\bibinfo {year} {2024})},\
  \Eprint {https://arxiv.org/abs/2311.01671} {arXiv:2311.01671 [hep-th]}
  \BibitemShut {NoStop}%
\end{thebibliography}%

\end{document}